\documentclass[reprint,
 amsmath,amssymb,
prb
]{revtex4-2}

\usepackage{graphicx}
\usepackage{dcolumn}
\usepackage{bm}

\usepackage[svgnames]{xcolor}
\usepackage{subfig}
\usepackage{subcaption}
\usepackage{mathrsfs}
\usepackage{tikz}
\newcommand{\figref}[1]{FIG.\ref{#1}}

\definecolor{myblue}{rgb}{0.07, 0.62, 1.00} 
\definecolor{mycyan}{rgb}{0.10, 0.90, 0.90}
\definecolor{mymagenta}{rgb}{1 0 1}

\begin{document}

\title{Quantum vortex driven Kelvin wave in the thermal background of Superfluid Helium}%

\author{Simone Scollo$^{1,2}$}
\email{simone.scollo@oca.eu}
\author{Luca Galantucci$^{3}$}
\author{Giorgio Krstulovic$^{2}$}

\affiliation{$^{1}$Universit\'e C\^ote d'Azur, Observatoire de la C\^ote d'Azur, CNRS, Laboratoire
J. L. Lagrange, Boulevard de l'Observatoire CS 34229 - F 06304 NICE Cedex 4, France}
\affiliation{$^{2}$
Universit\'{e} C\^{o}te d'Azur, CNRS, Institut de Physique de Nice (INPHYNI), 17 rue Julien Lauprêtre, 06200 Nice, France}
\affiliation{$^{3}$Istituto per le Applicazioni del Calcolo M. Picone, IAC-CNR, Via dei Taurini 19, Roma 00185, Italy}

\begin{abstract}
We present numerical evidence that Kelvin waves (KWs) on quantized vortices in superfluid helium can be directly observed in the normal fluid component at finite temperatures. Using the  Fully cOUpled loCAl model of sUperfLuid Turbulence (FOUCAULT) model, we analyze the propagation and temperature dependence of KWs by simultaneously measuring the dispersion of waves on the vortex displacement and the normal fluid velocity. 
The results demonstrate that the normal fluid supports a coherent KW-like response, with a dispersion relation matching that of the vortex filament (VF). Unlike the Schwarz model where there is almost no temperature dependence, in FOUCAULT KWs frequency and damping both depend on temperature, highlighting the role of mutual friction in mediating the coupling between the two fluids. These findings open a pathway for experimental observation of KWs in the normal phase using tracer based visualization.
\end{abstract}

\maketitle

\section{Introduction}

When cooled below the lambda point (approximately at $T_{\lambda}=2.17 \, \rm K$ at saturated vapor pressure) $^4$He, undergoes a phase transition to a superfluid state known as helium II \cite{Kapitza1938,Allen1938}. The dynamics of this phase is significantly determined by quantum mechanical phenomena, leading to a number of extraordinary properties. In fact, in pure superfluid state (in the absolute zero temperature limit), helium~II flows without carrying any entropy and without any measurable viscosity, allowing it to move through narrow channels or porous materials with no resistance and no heat transfer. Furthermore, the vorticity of the superfluid flow is entirely confined to topological defects of the superfluid order parameter, of atomic core size ($a_0\approx10^{-8}\,\rm{cm}$). These defects are effectively one-dimensional objects \cite{Onsager1949,Feynman1955} (for that usually referred to as \textit{vortex-lines}) around which the circulation of the velocity field is quantized in terms of the quantum of circulation $\kappa = h/m_4$, where $h$ is Planck's constant and $m_4$ the mass of a helium atom.

The dynamics of helium~II can be described employing a phenomenological framework introduced by Tisza and Landau \cite{Tisza1938,Landau1941}, \textit{the two-fluid model}. The latter describes helium II as an intimate mixture of two interpenetrating and inseparable fluid components: a \textit{superfluid} inviscid component, loosely related to the phenomenon of Bose-Einstein Condensation, with density $\rho_s$ and momentum density $\rho_s \mathbf{v}_s$; a \textit{normal} fluid viscous component, akin to a Navier-Stokes classical fluid, with density $\rho_n$ and momentum density $\rho_n \mathbf{v}_n$. The total density of helium~II is $\rho=\rho_s+\rho_n$ and its total momentum density $\mathbf{j}=\rho_s \mathbf{v}_s + \rho_n \mathbf{v}_n$. The relative density weight of each component $\rho_s/\rho_n$ is temperature-dependent, with the superfluid fraction increasing as temperature decreases and with the normal fraction exhibiting the opposite behavior. These two fluid components do not interact unless quantum vortices are nucleated  via \textit{e.g.} the rotation of helium~II, a sufficiently rapid temperature quench through the lambda transition, the mechanical or thermal stirring of helium~II. The presence of vortices couples the two fluid components via the mutual friction force \cite{Hall1956a,Hall1956b,barenghi1983friction,Donnelly1991}, proportional to the relative velocity between vortices and the normal fluid, which may trigger an additional energy dissipation on top of the viscous dissipation arising from internal frictions in the normal fluid.

This mutual friction force plays a crucial role in determining the characteristics of the collective motion of discrete sets of vortices, a dynamical state known as quantum turbulence \cite{barenghi-etal-2023,Barenghi_Skrbek_Sreenivasan_2023}, and in dictating the dynamics of individual vortices. Regarding the former, in the circumstance where the flow is stirred mechanically, the coupling of the flow components at scales larger than the average inter-vortex spacing determines the emergence of a classical Kolmogorov energy spectra $E(k) \sim k^{-5/3}$ \cite{barenghi-etal-spectra-2014}, $k$ being the wavenumber magnitude, for both the superfluid and the normal fluid. On the opposite, in flows generated thermically such as thermal counterflows \cite{vinen-1957c}, mutual friction is responsible for the observation of non-classical normal fluid energy spectra \cite{gao-etal-2017} and non-classical statistical distributions of normal velocity components \cite{galantucci-etal-2026}. In the context of single (or few) vortex dynamics, by governing the energy exchange between vortices and normal fluid, the mutual friction force determines the characteristics of vortex reconnections \cite{Stasiak2025a,Stasiak2025b} and the dynamics of excitations on the vortex line themselves, namely helical perturbations of the vortex line around its equilibrium position known as Kelvin waves (KWs). Depending on the features of the normal fluid flow, KWs may undergo growth or damping \cite{glaberson1974instability,Stasiak2024} by gaining or injecting energy in the normal fluid, respectively. The interplay between KWs on vortex filaments and normal fluid dynamics is precisely the topic of the present research.  

Theoretically, KWs were discovered by Lord Kelvin \cite{Kelvin1880} investigating the helical perturbations of a vortex line in an inviscid fluid whose dynamics is prescribed by the incompressible Euler equation. In particular, KWs are circularly polarized waves that propagate along the vortex line with a dispersion relation given, in the long wavelength limit ($ka_0 \ll 1$), by
\begin{equation}
\omega_k = -\frac{\kappa k^2}{4\pi} \left[ \ln\left(\frac{1}{ka_0}\right) + c \right],
\label{eq:KW_disp} 
\end{equation}
where $\omega_k$ is the angular frequency of the wave, $k$ is the wavenumber along the vortex line and $c$ is a constant of order unity that depends weakly on the vortex geometry. For instance, for a hollow vortex core $c = \ln 2 - \gamma$, with $\gamma=.5772$ the Euler-Mascheroni constant \cite{Roberts2003}.

In superfluid helium, KWs play a crucial role in the dynamics of quantum turbulence, particularly in the transfer of energy across scales and the decay of turbulence at very low temperatures \cite{Vinen2001,Kivotides2001,Leadbeater2003,kozik-svistunov-2004,vinen2003kelvin}, prompting several theoretical investigations \cite{Lvov2010,Krstulovic2012,Baggaley2014,Clark2015,Muller2020}. From the experimental point of view, studies on KWs have been performed in helium II in recent years \cite{Fonda2014,peretti-etal-2023,Minowa2025} thanks to the development of innovative visualization techniques, employing silicon nanoparticles \cite{minowa-etal-2022} or solid hydrogen/deuterium micron-sized particles \cite{Bewley2006,guo2014visualization, vessaire2025cryogeniclagrangianexplorationmodule} which interact with both the normal fluid and quantum vortices \cite{svancara-etal-2021} and may get trapped onto the latter. In this last circumstance, particles decorate vortices allowing the observation of their dynamics, including its three-dimensional reconstruction \cite{Minowa2025}, which could potentially lead to a detailed characterization of KWs dynamics
given the degree of control and reproducibility recently achieved \cite{peretti-etal-2023}. 

Recent insight about the interaction between KWs and particles in superfluid helium at very low temperatures has been provided by the numerical simulations using the Gross-Pitaevskii equation \cite{Giuriato2019,Giuriato2020}. This approach is however limited to very low temperatures, where the normal fluid component is almost absent, and to flow length scales which are too small compared to typical experimentally results.

In order to make direct comparison with ongoing experiments \cite{peretti-etal-2023}, in the present work we employ a recently developed numerical model named FOUCAULT (Fully cOUpled loCAl model of sUperfLuid Turbulence) \cite{Galantucci2020}, which accurately captures the coupled dynamics of the superfluid vortices and the normal fluid across a wide range of working temperatures achievable experimentally where the normal fraction is non negligible ($T > 1.5$K) and probing the flow at length scales much larger than the vortex core at scales experimentally accessible. Compared to past frameworks, \textit{e.g.} the HVBK model \cite{Hall1956a,Hall1956b,bekarevich-khalatnikov-1961,nemirovskii-fiszdon-1995} and the the pioneering model introduced by Schwarz \cite{schwarz-1978}, which provided a foundational description of vortex-filament dynamics at finite temperature, FOUCAULT takes into account the singular nature of quantum vortices describing them as filaments of infinitesimal length (neglected in the HVBK equations) and self-consistently couples the motion of quantum vortices to that of the normal fluid, each influencing the dynamics of the other (mutual interaction absent in Schwarz's equations of motion). FOUCAULT has been successfully employed in past studies to investigate various aspects of quantum turbulence, including vortex reconnections, energy spectra, and the interaction between vortices and normal fluid eddies \cite{Galantucci2023,Stasiak2024,Stasiak2025a,Stasiak2025b,galantucci-etal-2026}. Remarkably, among all available numerical models FOUCAULT is able to predict the observed decay of vortex rings with the best accuracy \cite{Tang2023}.

In this work, FOUCAULT will be used to study the dynamics of KWs in superfluid helium at finite temperatures, where the role played by the interaction with the normal component is of fundamental importance. We will focus on the manifestation of KWs linear response in the normal fluid component when the system is fully coupled and compare the results with the case without back reaction using the Schwarz model. We will show that at finite temperatures it is possible to observe the propagation of KW-like perturbation on the normal fluid (induced by a direct forcing of the quantum vortex displacements) with a spatio-temporal signature that is temperature dependent and coherent with the dispersion directly measured on quantum vortices.

This paper is organized as follows: in section II we briefly summarize the framework used to simulate the dynamics of superfluid helium at finite temperatures, both the Schwarz and the FOUCAULT models, with a discussion of the properties of KWs in quantum turbulence and their expected behavior at finite temperatures. In section III we describe details and parameters of the numerical methods employed to obtain our results described in section IV. Finally, in section V we summarize our findings and discuss their implications for future experimental studies.

\begin{figure}[ht!]
    \centering
    \includegraphics[width=0.5\textwidth]{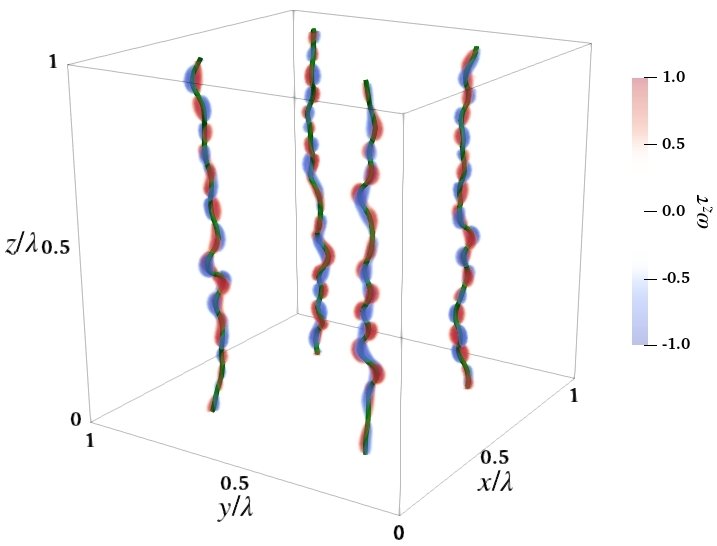}
    \caption{Visualization of the vortex lines configuration (in green) and the normal fluid z vorticity $\omega_z$ (in blue/red) at $T=1.95 \, \rm K$ at a given time. The vortex lines are perturbed by an ensemble of KWs with independent random phases, leading to a dipolar structure in the normal fluid vorticity around the vortex core.
    In this simulation the domain is a cube of size $\lambda$, and the amplitude of the waves is exaggerated to make the disturbance visible, routhly $50A_0=0.05\lambda$.}
    \label{fig:Fig1}
\end{figure}

\section{Modeling finite temperature superfluid helium}
\label{sec:modeling_he_ii}
\subsection{FOUCAULT}
To study numerically the dynamics of helium II at finite temperatures, we employ the fully coupled model FOUCAULT \cite{Galantucci2020}. Vortex lines are hence safely parametrized as one-dimensional space curves $\mathbf{s}(\xi,t)$, $\xi$ and $t$ being arclength and time respectively, and discretion with line elements of typical size $\delta$ such that $a_0 \ll \delta \ll \ell$. The Lagrangian equations of motion of the vortex lines are obtained by balancing the mutual friction force with the Magnus force leading to the following 
\begin{equation}
\begin{aligned}
\dot{\mathbf s}
  =\mathbf v_s(\mathbf s) &+ \frac{\rho_n}{\rho}\left [ \bigl(\mathbf v_n(\mathbf s) - \mathbf v_s(\mathbf s)\bigr)\cdot\mathbf{s}'\right ]\mathbf{s}' \\
    &+ \beta\,\mathbf s' \times \bigl(\mathbf v_n(\mathbf s) - \mathbf v_s(\mathbf s)\bigr) \\
    &+ \beta'\,\mathbf s' \times \left[\,\mathbf s' \times \bigl(\mathbf v_n(\mathbf s) - \mathbf v_s(\mathbf s)\bigr)\right]\,\, ,
    \label{eq:s_dot_Fclt}
\end{aligned}
\end{equation}
where $\mathbf s' = \partial \mathbf s / \partial \xi$ is the unit tangent vector to the filament, $\mathbf v_s(\mathbf s)$ and $\mathbf v_n(\mathbf s)$ are the superfluid and normal fluid velocities at the point $\mathbf s$ and $\beta$, $\beta'$ are temperature and Reynolds number dependent mutual friction coefficients.

In this model the superfluid velocity $\mathbf v_s$ is given by the Biot-Savart law
\begin{equation}
\mathbf v_s(\mathbf x,t)
  = \frac{\kappa}{4\pi}
      \oint_{\mathcal L}
        \frac{\mathbf s'(\xi,t)\times\left(\mathbf x - \mathbf s(\xi,t)\right)}
             {\left|\mathbf x - \mathbf s(\xi,t)\right|^3}
        \;\mathrm d\xi \,\, ,
        \label{eq:Biort-Savart}
\end{equation}
$\mathcal L$ being the whole vortex line configuration. At the length scales probed both the superfluid and the normal fluid flows are incompressible. As the smallest scales investigated are much larger than $\lambda_{\rm MFR}$, the normal fluid velocity field $\mathbf v_n$ is governed by the incompressible Navier-Stokes equations with an additional mutual friction force term
\begin{equation}
\begin{aligned}
\frac{\partial \mathbf v_n}{\partial t} + (\mathbf v_n \cdot \nabla)\,\mathbf v_n 
  &= -\nabla \left( \frac{p}{\rho_n} \right)
    + \nu_n \,\nabla^2 \mathbf v_n \\
  &\quad + \frac{1}{\rho_n} \oint_{\mathcal L} \delta(\mathbf x - \mathbf s)\;\mathbf f_{ns}(\mathbf s)\;\mathrm d\xi
  \label{eq:NS_eq}
\end{aligned}
\end{equation}
\begin{equation}
\nabla \cdot \mathbf v_n = 0
\label{eq:div_free}
\end{equation}
where  $\nu_n$ is the kinematic viscosity of the normal fluid, $p$ is the pressure and $\mathbf f_{ns}(\mathbf s)$ is the mutual friction force per unit length exerted by the vortex line on the normal fluid at the point $\mathbf s$, given by \cite{Galantucci2015}
\begin{equation}
\begin{aligned}
\mathbf f_{ns}(\mathbf s)
  = -\rho_n \kappa \,\mathbf s' \times\bigl(\dot{\mathbf s} - \mathbf v_n\bigr)
    - \,D\,\mathbf s' \times \bigl[\,\mathbf s' \times (\dot{\mathbf s} - \mathbf v_n) \bigr]
    \label{eq:friction_force}
\end{aligned}
\end{equation}
where the viscous coefficient $D$ is determined by a classical low Reynolds number approach \cite{proudman-pearson-1957} and reads as follows
\begin{equation}
    D=\frac{4\pi\rho_n\nu_n}{\left[\frac{1}{2}-\gamma-\ln\left(\frac{|\mathbf v_{n_\perp}-\dot{\mathbf s}|a_0}{4\nu_n}\right)\right]},
\end{equation}
$\mathbf v_{n_\perp}$ being the normal fluid velocity projection on a plane orthogonal to $\mathbf s'$. More detail on the model and numerical implementation can be found in \cite{Galantucci2020}.

\subsection{Schwarz model}
Simulations performed within the FOUCAULT framework will be compared to simulations realized with the pioneering Schwarz model \cite{schwarz-1978,Schwarz1985,Schwarz1988}, where the back-reaction of the vortex motion on the normal fluid is neglected (the latter being hence prescribed a priori, determined by the boundary conditions of the flow) and the mutual friction coefficients considered coincide with the friction parameters determined by Hall and Vinen \cite{Hall1956a,Hall1956b}. The consequent Lagrangian equations of motion for a vortex line reads as follows
\begin{equation}
\begin{aligned}
\dot{\mathbf s}
  = \mathbf v_{s}(\mathbf s)
    &+ \alpha\,\mathbf s' \times \bigl(\mathbf v_n(\mathbf s) - \mathbf v_s(\mathbf s)\bigr) \\
    &- \alpha'\,\mathbf s' \times \left[\,\mathbf s' \times \bigl(\mathbf v_n(\mathbf s) - \mathbf v_s(\mathbf s)\bigr)\right],
\end{aligned}
\label{eq:Schwarz_model}
\end{equation}
where $\alpha$ and $\alpha'$ are the mutual friction coefficients. In the simulations performed by Schwarz \cite{schwarz-1978,Schwarz1985,Schwarz1988} the superfluid velocity $\mathbf v_s$ was approximated employing the Local Induction Approximation (LIA), where each vortex line element is only advected by the local contribution, leading to a simpler expression for the superfluid velocity
\begin{equation}
  \mathbf v^{\rm LIA}_{s} = \Lambda \,\mathbf s' \times \mathbf s'',
  \label{eq:LIA}
\end{equation}
where $\Lambda = (\kappa/4\pi)\,\ln(\ell/a_0)$, $\ell$ is the inter-vortex distance and $\mathbf s''=\partial^2 \mathbf s/\partial \xi^2$ is the local curvature vector. In our numerical simulations performed with the Schwarz model we will instead compute the whole (suitably de-singularized \cite{baggaley-barenghi-2011c}) Biot-Savart integral reported in  Eq.~(\ref{eq:Biort-Savart}). In our simulations employing the Schwarz model the normal fluid velocity is imposed $\mathbf v_n\equiv 0$.

\subsection{Kelvin waves at finite temperature}
\label{subsec:KW_finite_T}
At finite temperatures, the presence of the normal fluid component and the associated mutual friction modify the dynamics of KWs. In the framework of the Schwarz model described by Eq.~(\ref{eq:Schwarz_model}), the mutual friction introduces a damping term and a small temperature dependent frequency attenuation that affects the propagation of KWs. If $\omega_k$ denotes the Kelvin wave dispersion relation at zero temperature (Eq.~\eqref{eq:KW_disp} for $a_0k\ll1$), at finite temperature Eq.~(\ref{eq:Schwarz_model}) predicts the temperature-modified dispersion relation $\omega_k^{T}$  that reads \cite{Krstulovic2023}
\begin{equation}
\omega_k^{T} = (1-\alpha')\omega_k \, ,
\label{eq:KW_disp_termal}
\end{equation}
and the amplitude $S(t)$ follows the exponential damping 
\begin{equation}
S(t) = S_0 e^{-\sigma_k t} \, ,
\label{eq:KW_damping_termal}
\end{equation} 
where $\sigma_k = \alpha \omega_k$, $S_0$ being the initial amplitude. Thus, the dispersion relation of KWs at finite temperatures is only scaled by a factor $(1-\alpha')$ with respect to the zero-temperature limit. However, since $(1-\alpha')\approx 1$, the temperature dependence of the dispersion relation is expected to be weak in the Schwarz model. 

In the FOUCAULT framework, the dynamics is more complex as a KW induces a localized perturbation in the normal fluid. To understand the normal fluid velocity response to KWs, we consider small perturbations of a straight vortex $\mathbf s(\xi,t)=(X(\xi,t), Y(\xi,t), \xi)$, corresponding to a superposition of random small amplitude KWs that reads explicitly 
\begin{equation}
    s(\xi,t) = \sum_{q_z} s^0_{q_z} e^{i(q_z \xi - \omega_{q_z}t)},\label{eq:KWsFOUCAULT}
\end{equation}
where $s(\xi,t) =X(\xi,t)+i Y(\xi,t)$.

We then look at the leading order contribution of the mutual friction to the normal fluid component, which in complex variables reads
\begin{equation}
\begin{aligned}
    &\mathscr{L}_{\rm NS}V_n=-(\partial_x+i\partial_y)\left(\frac{p}{\rho_n}\right)\\
    &-\oint_{\mathcal L} \delta(\mathbf x-\mathbf s) (i\kappa-D)(\dot s + \partial_\xi s \,w_n - V_n) d\xi,
\end{aligned}
\end{equation}
with $V_n=u_n+iv_n$ and where $u_n, v_n, w_n$ are the normal fluid velocity along the $x,y,z$ respectively. The operator $\mathscr{L}_{\rm NS}$ is given by the following expression $\mathscr{L}_{\rm NS}=\partial_t + \mathbf v_n \cdot \nabla - \nu_n \nabla^2$.

In the equilibrium state, the vortex line is straight and aligned along the $z$-axis, implying that in the linear response limit $w_n$ is negligible with respect to $u_n$ and $v_n$.

Using the explicit form of the perturbation we obtain
\begin{equation}
\begin{aligned}
    &\mathscr{L}_{\rm NS}V_n=-(\partial_x+i\partial_y)\left(\frac{p}{\rho_n}\right)+\mathscr{I}(V_n)\\
    &-\sum_{q_z}s^0_{q_z}\oint_{\mathcal L} \delta(\mathbf x-\mathbf s) \left[(\kappa+i D)\omega_{q_z}e^{i(q_z \xi - \omega_{q_z} t)}\right] d\xi,
\end{aligned}
\end{equation}
where $\mathscr{I}(V_n)$ is an abbreviation for the linear integral in the preceding formula.

We are interested to isolate the forcing term due to the KWs perturbation, so we Fourier transform the equation in space, obtaining for the forcing term the following expression
\begin{equation}
    -\sum_{q_z} s^0_{q_z} e^{-i\omega_{q_z} t} (\kappa+i D)\omega_{q_z}\oint_{\mathcal L}e^{iq_z\xi}e^{-i\mathbf{k}\cdot \mathbf{s}(\xi)} d\xi,
\end{equation}
where the integrand can be expanded as $e^{-i\mathbf{k}\cdot \mathbf{s}(\xi)}=e^{-i k_z \xi}e^{-i\mathbf{k}_\perp\cdot \mathbf{s}_\perp}\approx e^{-ik_z \xi}\left[1-i\mathbf{k}_\perp\cdot \mathbf{s}_\perp\right]$ such that at leading order (for small perturbations) we have
\begin{widetext}
\begin{equation}
\begin{aligned}
    \widehat{\mathscr{L}_{\rm NS}V_n}+(i k_x - k_y)\left(\frac{\widehat{p}_k}{\rho_n}\right)-\widehat{\mathscr{I}(V_n)}\approx&-\sum_{q_z} s^0_{q_z} (\kappa+i D)\omega_{q_z}e^{-i\omega_{q_z} t}\oint_{\mathcal L}e^{i(q_z-k_z)\xi} d\xi\\
    =&-\sum_{q_z} s^0_{q_z} (\kappa+i D)\omega_{q_z}e^{-i\omega_{q_z} t}2\pi\delta_{k_z,q_z}\,= -s^0_{k_z} (\kappa+i D)\omega_{k_z}e^{-i\omega_{k_z} t}2\pi.
    \label{eq:KW_forcing_general}
\end{aligned}
\end{equation}
\end{widetext}
This result tells us that the normal fluid velocity is directly forced by KWs and therefore should exhibit a spatio-temporal signature similar to that of the KWs on the vortex line.

\section{Numerical Treatment}
 
\subsection{Simulation parameters}
The numerical simulations are performed in a periodic box of size $L_x=L_y=2\pi$ and $L_z=16\pi$. The normal fluid is discretized on a uniform Cartesian grid with $n_x=n_y=128$ and $n_z=1024$ points using a pseudo-spectral method with a second-order Runge-Kutta scheme for time integration. Quantum vortices are discretized into a series of points along its length, with a spatial resolution $\delta\approx\Delta z$. The Biot-Savart integral Eq.~(\ref{eq:Biort-Savart}) is fully computed using a splitting regularization on local and nonlocal contribution \cite{baggaley-barenghi-2011c}. The vortex line equations of motion (Eq.~\ref{eq:s_dot_Fclt} or Eq.~\ref{eq:Schwarz_model}, depending on the model employed) is evolved in time using a second-order Runge-Kutta scheme with a timestep $\Delta t_{VF}$, which can be smaller than the normal fluid timestep $\Delta t_{NS}=0.04$ to ensure numerical stability.

We have considered three different finite temperature values, $T=1.7 \, {\rm K}$, $T=1.95 \, {\rm K}$, and $T=2.1 \, {\rm K}$, corresponding to different values of the mutual friction coefficients ($\beta$ and $\beta'$ in the FOUCAULT model, $\alpha$ and $\alpha'$ in the Schwarz model), 
as well as different normal fluid densities $\rho_n$ and viscosity $\nu_n$. The relevant physical parameters used in the simulations are summarized in Table \ref{tab:table1}. We have also performed simulations at $T=0$, where only the equations of motion of vortex filament are integrated (in the zero-temperature limit, both models reduce to the same equation $\dot{\mathbf s}  = \mathbf v_{s}(\mathbf s)$).

\begin{table}[h!]\caption{\label{tab:table1}%
  Numerical value of the physical parameters used in the simulations at different temperatures. Values taken from \cite{Donnelly1998}.
}
\begin{ruledtabular}
\begin{tabular}{lcccccr}
&\textrm{T(K)}&
\textrm{$\Gamma=\kappa/\nu_n$}&
\textrm{$\rho_{ns}=\rho_n/\rho_s$}&
\textrm{$\alpha$}&
\textrm{$\alpha'$}\\
\colrule
\rule{0pt}{1.25em}
&0 & - & 0 & 0 & 0 & \\
&1.7 & 2.565 & 0.2958 & 0.126 & 0.012 &\\
&1.95 & 5.005 & 0.9298 & 0.237 & 0.011 &\\
&2.1 & 5.9736 & 2.8467 & 0.479 & -0.024 &\\
\end{tabular}
\end{ruledtabular}
\end{table}

The equations used are in dimensionless units, which are related to physical units by fixing the dimensional length scale $\lambda$ such that the dimensional timescale $\tau=\lambda^2\nu^0/\nu_n$, where $\nu^0=0.005$ is the set numerical viscosity and $\lambda=1.59\times 10^{-2} \, \rm cm$, corresponding to an experimental length scale in the x-direction of $0.1 \, \rm cm$. These values lead to the following range of timescale $\tau$: $2.99\times10^{-1} \, \rm s$ at $T=2.1 \, \rm K$ to $1.29\times10^{-1} \, \rm s$ at $T=1.7 \, \rm K$ (The case $T=0 \, {\rm K}$ is chosen to have the same timescale of $T=1.7 \, {\rm K}$). In the following all the quantities will be expressed in these dimensionless units, rescaling the time at different temperatures to be consistent with $\tau(T=1.7 \, {\rm K})$, so for a temperature $T^*$ the dimensionless time is rescaled as  $\tau(T=1.7 \, {\rm K})/\tau(T^*)$.

To properly resolve the dispersion relation of KWs for small wavenumber, the computational domain must be sufficiently elongated along the vortex line direction (z-axis). This is necessary to achieve a high spectral resolution in the wavenumber space, allowing for accurate measurements of the KW dispersion relation at low $k_z$ values. Tacking $L_z>>1$ implies that the lowest frequency $\omega_{\rm min}<<1$, which results in very large period $T_{\rm max}>>1$. However, the simulation duration is constrained by viscous damping. Consequently, low $k$ measurements are less reliable and cannot be improved within the limitations of the present code and computational resources.

\subsection{Initial conditions}
The VF initial condition consists of four vortex lines aligned along the $z$-axis oriented with opposite circulation with respect to its nearest neighbors to ensure zero net circulation in the box to preserve periodic boundary conditions. They are located in their equilibrium position and each is perturbed by an independent superposition of KWs with random phases
\begin{equation*}
    s_i(z,t)|_{t=0}=A_0\sum_{n=1}^{N} \frac{1}{\sqrt{k^n_z}}\left[e^{i\left(k^n_z z+\phi^+_{n,i}\right)}+e^{i\left(-k^n_z z+\phi^-_{n,i}\right)}\right]
\end{equation*}
where $k_z^{n}=n\,2\pi/Lz$, $\phi^\pm_{n,i}$ independent random phases, each uniformly distributed in  $[0,2\pi]$, $N=n_z/3$ is the number of modes considered and $A_0=0.001$ is the small initial amplitude (the same for all lines) to remain in the linear regime. In the Schwarz simulations the normal fluid is imposed to zero, while in the FOUCAULT simulations, the normal fluid is initially at rest and, fixing the position of the vortices for a transient, it is left to evolve with the coupling to relax on the stable configuration leading to the dipole structure shown in figure \ref{fig:Fig1} for a larger amplitude $A_0$ to make the perturbation visible. With this initial condition the simulations are run for a sufficient time in order to allow KW propagation and interaction with the normal fluid.

\subsection{Measurement of the dispersion relation}

To quantitatively assess the KW dynamics, we compute the spatio-temporal spectrum of the vortex line displacements and of the normal fluid velocity field. 

The vortex line position is reconstructed as a complex field
\begin{equation}
S(z,t) = X(z,t) + i\,Y(z,t),
\end{equation}
where $X$ and $Y$ are the transverse displacements of the vortex from its unperturbed position along the $z$-axis. The corresponding Fourier transform in both space and time,
\begin{equation}
\hat{S}(k_z,\omega) = \int\!\!\int S(z,t)\,e^{-i(k_z z - \omega t)} \, \mathrm{d}z\,\mathrm{d}t,
\label{eq:S_k_w}
\end{equation}
yields the spatio-temporal spectrum $|\hat{S}(k_z,\omega)|^2$, which directly reveals the dispersion relation $\omega(k_z)$ of propagating excitations. To reduce statistical noise and directly extract the dispersion relation, after averaging over the four different lines, we measure  $\omega$ at each wavenumber $k_z$ using the weighted average
\begin{equation}
\tilde{\omega} (k_z)= 
\frac{\displaystyle \int \omega\,|\hat{S}(k_z,\omega)|^{2}\,\mathrm{d}\omega}
{\displaystyle \int |\hat{S}(k_z,\omega)|^{2}\,\mathrm{d}\omega}.
\label{eq:omega2}
\end{equation}
This method is particularly robust in the presence of weak damping and finite sampling time. In the following, we will refer to $\tilde{\omega}^{Sc}$ and $\tilde{\omega}^{F}$ as the measured dispersion relation of KWs on the Schwarz model and the FOUCAULT model respectively.

The same procedure is applied to the normal fluid velocity $v_x$, allowing direct comparison between the KW dispersion on the filament and the response in the normal fluid. For that we compute the spatio-temporal spectrum of the normal fluid velocity component $v_x$, restrict to the wavenumbers $\mathbf{k}=(0,0,k_z)$ aligned with the vortex line, as
\begin{equation}
\hat{V}_x(k_z,\omega) = \int\!\!\int v_x(\mathbf{x},t)\,e^{-i(k_z z - \omega t)} \, \mathrm{d}\mathbf{x}\,\mathrm{d}t,
\label{eq:V_k_w}
\end{equation}
to have the spatio-temporal spectrum $|\hat{V}_x(k_z,\omega)|^2$. The choice made here to focus on the $v_x$ component and not on the complex transverse velocity $\hat{V}_n = \hat{u}_n + i\hat{v}_n$ does not change the results since we are in the linear regime and the two components are expected to have the same spatio-temporal signature.

\begin{figure}[h!t]
    \centering
    \includegraphics[width=1\linewidth]{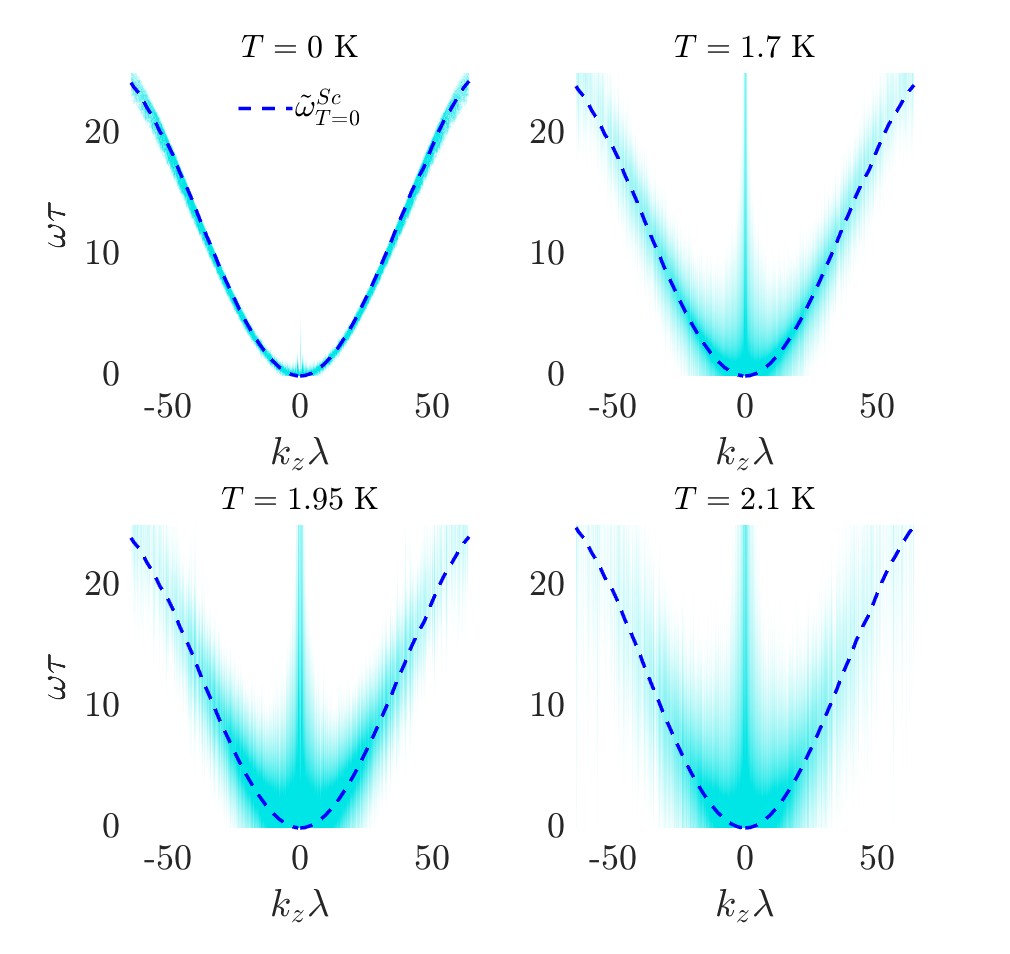}
    \caption{Blue heatmaps: dispersion relation of KWs on vortex lines simulated with the Schwarz model using the full Biot-Savart integral for 
    $T=0, \, 1.7, \, 1.95, \, 2.1 \, {\rm K}$ using Eq.~(\ref{eq:S_k_w}). The dispersion relation at temperature $T=0$ K calculated via Eq.~(\ref{eq:omega2}) is indicated (blue dashed line \textcolor{blue}{\rule[0.5ex]{0.2cm}{0.8pt} \rule[0.5ex]{0.2cm}{0.8pt}}) and superimposed on all cases.}
    \label{fig:Fig2}
\end{figure}

\section{Results and discussion}

\subsection{Kelvin waves in Schwarz model}
In order to reduce complexity and analyze separately the impact on the dispersion relation of the mutual friction on its own and of the mutual friction combined with the feedback of normal fluid fluctuations on the vortex itself, we first performed a simulation employing the Schwarz model, Eq.~(\ref{eq:Schwarz_model}), at $T=0, \, 1.7, \, 1.95, \, 2.1 \, {\rm K}$.

The blue heatmap plots in Figure~\ref{fig:Fig2} show the measured dispersion relation of KWs $|\hat{S}(k_z,\omega)|^2$ obtained employing Eq.~(\ref{eq:S_k_w}) at the different temperatures. Superimposed on the heatmaps we report in blue-dashed curves the $T=0$ dispersion relation $\tilde{\omega}^{Sc}_{T=0}$ determined using  Eq.~(\ref{eq:omega2}).

The results confirm the expected quadratic scaling of Eq.~(\ref{eq:KW_disp}) at low $k_z$, with a gradual increase of the thickness of the dispersion as temperature increases, corresponding to the enhanced mutual friction damping at increasing temperature described in the analytical results for small perturbations reported in Eq.~(\ref{eq:KW_damping_termal}). Importantly, Figure~\ref{fig:Fig2} clearly shows that the dispersion relation in the Schwarz model is almost temperature independent, consistently again with analytics, Eq.~(\ref{eq:KW_disp_termal}). The consistency between the measured and theoretical dispersion validates both the numerical implementation and the measurement method based on Eq.~(\ref{eq:omega2}).

\subsection{Kelvin waves in FOUCAULT model}

\begin{figure*}[h!t]
    \centering
    \subfloat[$T=1.7 \, {\rm K}$]{\includegraphics[width=0.3\linewidth]{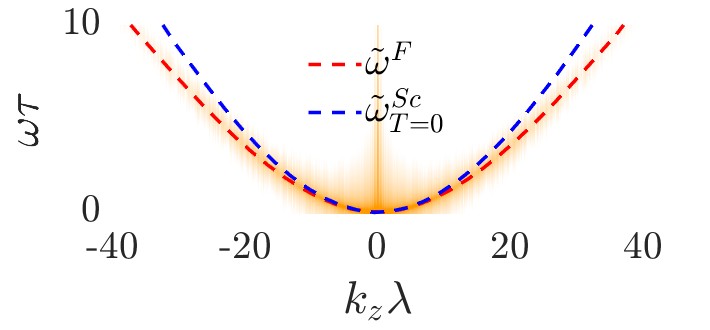}}
    \quad
    \subfloat[$T=1.95 \, {\rm K}$]{\includegraphics[width=0.3\linewidth]{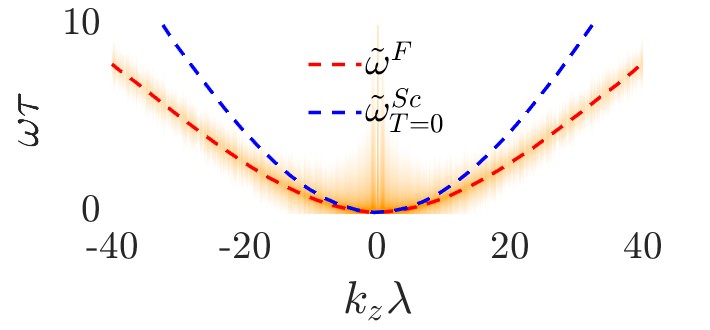}}
    \quad
    \subfloat[$T=2.1 \, {\rm K}$]{\includegraphics[width=0.3\linewidth]{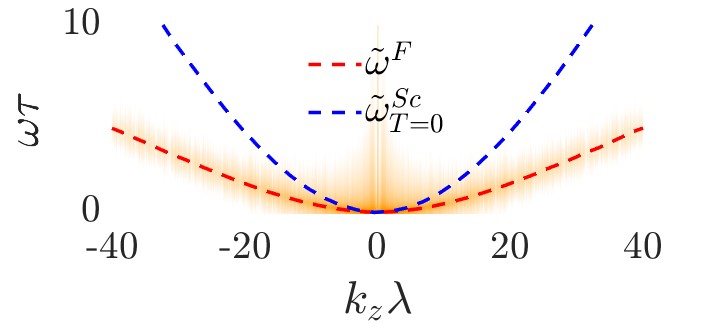}} \\
    \subfloat[$T=1.7 \, {\rm K}$]{\includegraphics[width=0.3\linewidth]{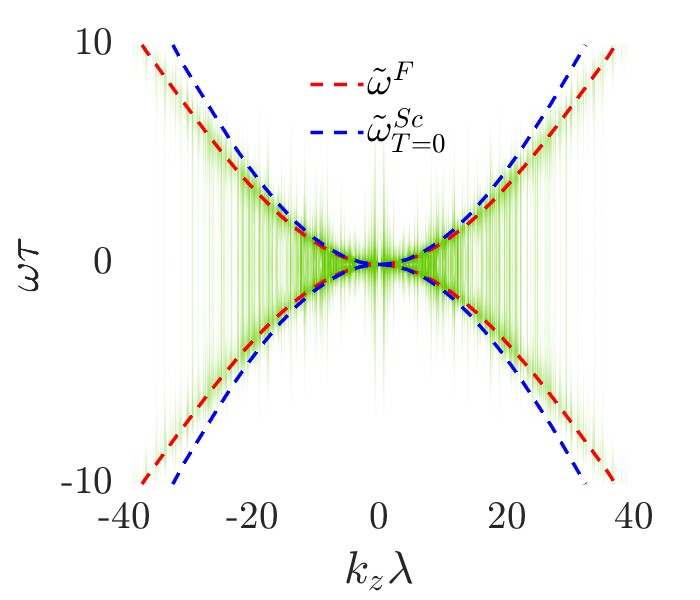}}
    \quad
    \subfloat[$T=1.95 \, {\rm K}$]{\includegraphics[width=0.3\linewidth]{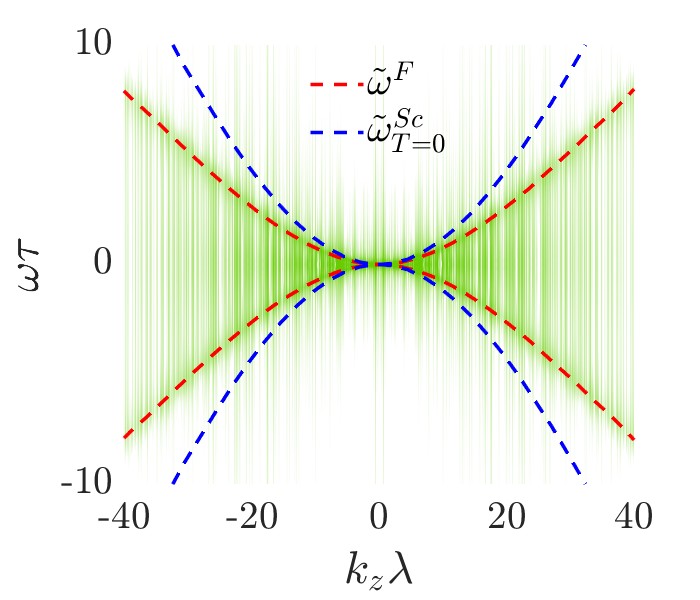}}
    \quad
    \subfloat[$T=2.1 \, {\rm K}$]{\includegraphics[width=0.3\linewidth]{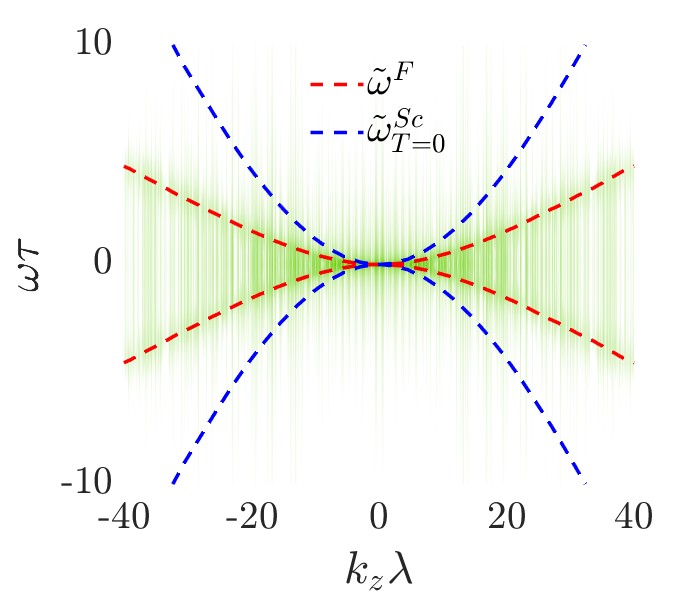}}
    \caption{Dispersion relation for temperatures $T=1.7 \, {\rm K}$ (a,d), $T=1.95 \, {\rm K}$ ((b,e) and $T=2.1 \, {\rm K}$ (c,f) determined 
    on the basis of FOUCAULT numerical simulations for: KWs on quantum vortices, $|\hat{S}(k_z,\omega)|^2$ (a,b,c, orange colored heatmaps); KWs in the normal fluid
    $|\hat{V}_x(k_z,\omega)|^2$ (d,e,f, green colored heatmaps). 
    We also show the dispersion relation of KWs on vortices computed with the Schwarz model at $T=0$ $\tilde{\omega}^{Sc}_{T=0}(k_z)$ 
    (blue dashed line \textcolor{blue}{\rule[0.5ex]{0.2cm}{0.8pt} \rule[0.5ex]{0.2cm}{0.8pt}}, as in \figref{fig:Fig2}).
    and the dispersion relation $\tilde{\omega}^F_T(k_z)$ computed with Foucault at corresponding temperatures
    (red dashed line \textcolor{red}{\rule[0.5ex]{0.2cm}{0.8pt} \rule[0.5ex]{0.2cm}{0.8pt}}).}
    \label{fig:Fig3}
\end{figure*}

We now turn to study the simulations of the fully coupled FOUCAULT model, which explicitly includes the back reaction of the vortex motion on the normal fluid and, conversely, the impact of normal fluid fluctuations on the dynamics of vortices. Figure~\ref{fig:Fig3} displays the spatio-temporal spectra of the vortex displacement $|\hat{S}(k_z,\omega)|^2$ (Fig.~\ref{fig:Fig3} (a-c) in orange colored heatmaps) and of the normal fluid velocity $|\hat{V}_x(k_z,\omega)|^2$ (Fig.~\ref{fig:Fig3} (d-f) in green colored heatmaps) for the three temperatures investigated. On top of the heatmaps we superimpose $\tilde{\omega}^{Sc}_{T=0}$ and $\tilde{\omega}^F_T$, the spatio-temporal weighted dispersion relation of KWs on vortices calculated from FOUCAULT simulations at $T\neq 0$. From Fig.~\ref{fig:Fig3} (a-c) it clearly emerges the novel feature that $\tilde{\omega}^F_T$ is strongly temperature dependent, in contrast to the results obtained with the Schwarz model.

In addition, as expected from our analytical approach reported in Section~\ref{subsec:KW_finite_T}, at all temperatures a clear KW branch emerges also in the normal fluid signal $|\hat{V}_x(k_z,\omega)|^2$ (Fig.~\ref{fig:Fig3} (d-f), in green colored heatmaps): the frequencies measured in the normal phase do coincide with the frequencies $\tilde{\omega}^F_T(k_z)$ of KWs on vortices (the red dashed lines) within numerical uncertainty, demonstrating that the normal fluid responds coherently to the oscillatory motion of the quantized vortices. A cross coherence analysis, not shown here, confirms this interpretation with an high correlation on the dispersion relation ridge showing a phase alignment between the vortex displacement and the normal-fluid response. At higher temperatures, the KW frequency in both fluids, decreases and the spectral peak broadens, indicating stronger damping due to the viscous and frictional coupling between the two components.

\begin{figure}
    \centering
    \includegraphics[width=1\linewidth]{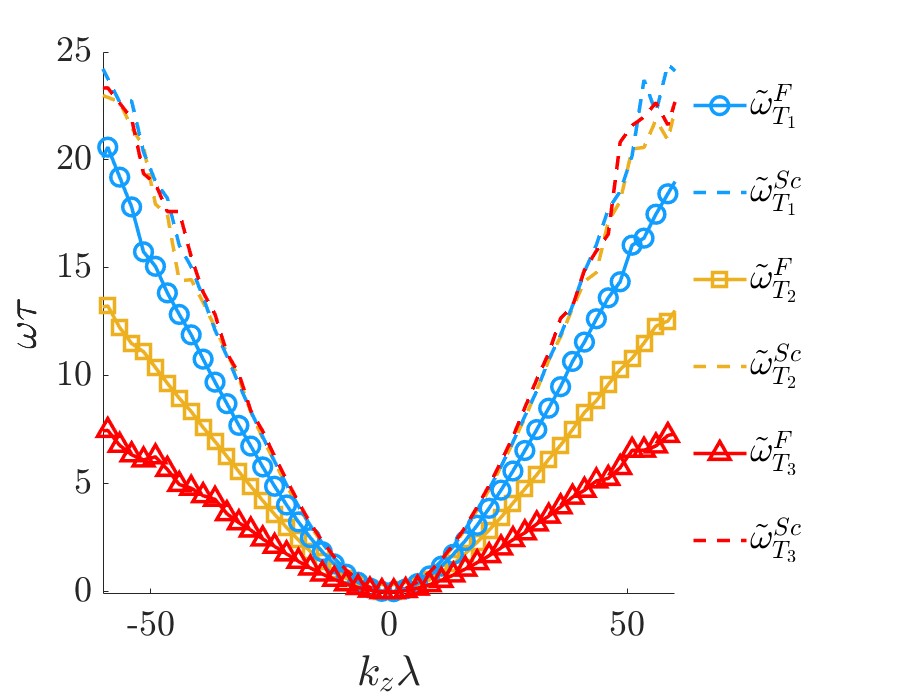}
    \caption{Dispersion relation of KWs on vortices $\tilde{\omega}^{Sc}_T(k_z)$ computed on the basis of the Schwarz model at the three working temperatures (dashed lines).
    Dispersion relation of KWs in the normal fluid $\tilde{\omega}^{F}_T(k_z)$ computed on the basis of the FOUCAULT model at the three working temperatures (solid lines with markers). Blue, yellow and red curves correspond to $T=1.7, \, 1.95, \, 2.1 \, {\rm K}$, respectively.}
    \label{fig:Fig4}
\end{figure}

The comparison between the results obtained with the Schwarz model and FOUCAULT are summarized in Fig.~\ref{fig:Fig4} which clearly illustrates that all results obtained from the Schwarz model collapse onto a single curve (as predicted analytically, Eq.~(\ref{eq:KW_disp_termal}), as $\alpha'\approx 0$ ), while, in contrast, the normal fluid measurements obtained from the FOUCAULT simulations exhibit a systematic temperature dependence: as $T$ increases, the KW branch in the normal fluid shifts to lower frequencies and becomes increasingly damped. This behavior directly reflects the enhanced energy exchange through mutual friction, which transfers part of the KW energy of vortices to the normal phase, then dissipated by viscosity.

These findings establish that KW are not confined to the inviscid superfluid component but can be observed as coherent oscillations in the normal fluid surrounding the vortex cores, making the dipolar structure of it to oscillate at the same frequency. This observation opens new perspectives for experimental detection of KW dynamics using normal fluid diagnostics, such as particle tracking velocimetry or tracer imaging, without requiring direct access to the superfluid core.

\subsection{Damping of Kelvin waves}

The damping of KWs is characterized by an exponential decay of wave amplitude over time with a rate $\sigma_k=\alpha \omega_k$ given by Eq.(\ref{eq:KW_damping_termal}) \cite{Krstulovic2023} stemming from the Schwarz framework in the linear approximation. This relation indicates that the damping rate is proportional to the wave frequency, with the proportionality constant set by the strength of mutual friction. In the fully coupled FOUCAULT model, the damping of KWs arises from a combination of mutual friction and viscous dissipation phenomena. The effective damping rate $\sigma_k$ may thus deviate from the simple Schwarz prediction, due to the complex interplay between these two mechanisms. 

\begin{figure}[ht!]
  \centering
  \includegraphics[width=1\linewidth]{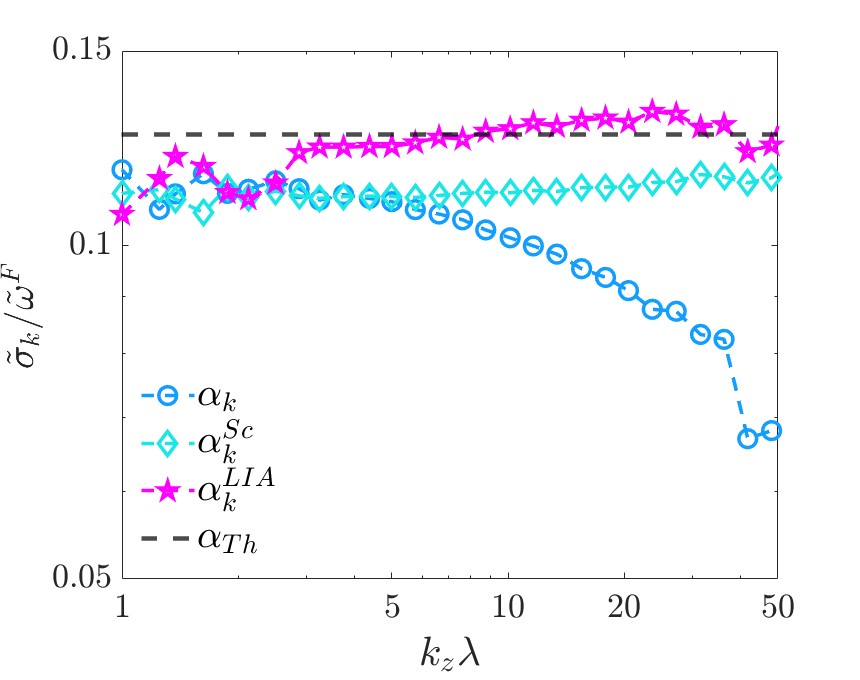}
  \caption{Measurement of the dimensionless damping rate $\alpha_k$ at $T=1.7 \, {\rm K}$ for the FOUCAULT model (blue circles \textcolor{myblue}{$\protect\tikz \protect\draw[very thick] (0,0) circle (0.1cm);$}), Schwarz model (cyan diamond \textcolor{mycyan}{\small $\mathbf{\Diamond}$}) and LIA model (magenta star \textcolor{mymagenta}{$\bigstar$}). The dashed black line represent the theoretical value $\alpha=0.126$.}
  \label{fig:Fig5}
\end{figure}

In our analysis, we directly measure $\tilde{\sigma}_k$ by tracking the temporal relaxation of the wave action, defined as $n(k_z,t)=<|\hat{S}(k_z,t)|^2>$, (where the average is performed over the four filaments), and extracting the damping rate $\alpha_k$ for each accessible wavenumber $k_z$ assuming the following exponential decay  
\begin{equation}
n(k_z,t) \sim e^{-2\tilde{\sigma}_k t}.
\end{equation}
Using the measured $\tilde{\omega}(k_z)$, we define the dimensionless ratio $\alpha_k \equiv \tilde{\sigma}_k/\tilde{\omega}(k_z)$ that quantifies the effective damping relative to the wavenumber $k_z$ and encapsulates the relative strength of dissipation to wave propagation. The calculation of $\alpha_k$ only depends on the vortex displacements. In order to compare the effective damping for the different models employed, we set the temperature to $T=1.7 \, {\rm K}$, and we calculate $\alpha_k$ from numerical simulations performed with FOUCAULT and the Schwarz model. We also compute $\alpha_k$ by evolving quantum vortices in the Schwarz model using the LIA, Eq.(\ref{eq:LIA}), approximation instead of the full Biot-Savart integral, Eq.(\ref{eq:Biort-Savart}), for the evaluation of the superfluid velocity on vortices. The resulting behaviors of $\alpha_k$ are shown in Fig.~\ref{fig:Fig5} together with the constant theoretical prediction $\alpha_k=\alpha=0.126$. As expected LIA gives the correct value showing that the measurement procedure is consistent for large enough $k_z$ (the theoretical prediction $\sigma_k=\alpha \omega_k$ is indeed obtained with LIA). The Schwarz model shows a constant value of $\alpha_k$ in the uncertainty, with a deviation from the theoretical value. This difference can be attributed to the finite amplitude of the initial perturbation, which introduces nonlinear effects not captured in the linear theory. The main result emerging from Fig.~\ref{fig:Fig5} is that the FOUCAULT model shows a trend deviating from the constant value returned by the Schwarz model, with $\alpha_k$ decreasing at increasing $k_z$. This behavior suggests that the back reaction of the normal fluid on the vortex dynamics introduces a scale-dependent modification to the KW damping, likely due to the complex interplay between mutual friction and viscous dissipation in the normal phase.

\begin{figure}[ht!]
  \centering
  \subfloat[$T=1.7 \, {\rm K}$]{\includegraphics[width=0.48\linewidth]{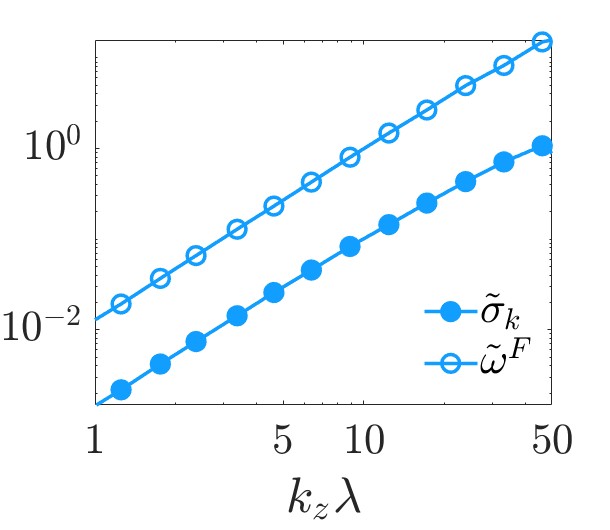}}\quad
  \subfloat[$T=1.95 \, {\rm K}$]{\includegraphics[width=0.48\linewidth]{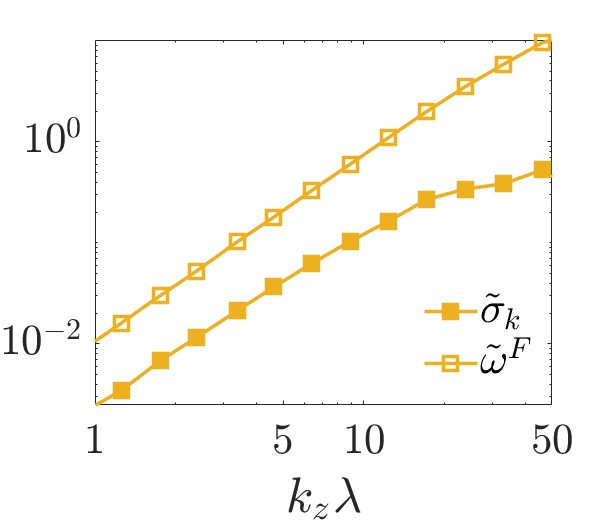}}\\
  \subfloat[$T=2.1 \, {\rm K}$]{\includegraphics[width=0.48\linewidth]{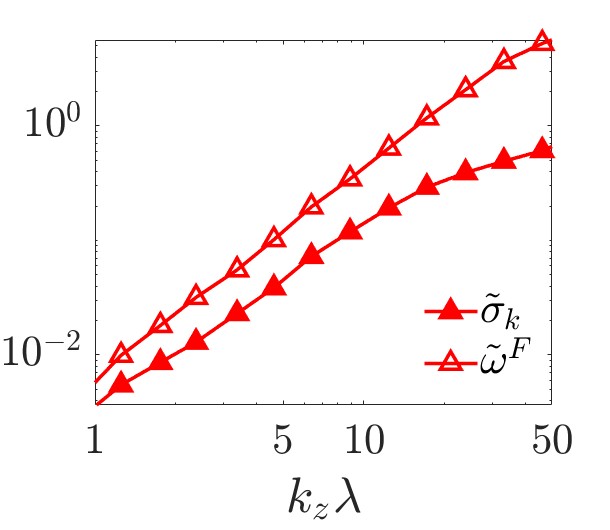}}\quad
  \subfloat[]{\includegraphics[width=0.48\linewidth]{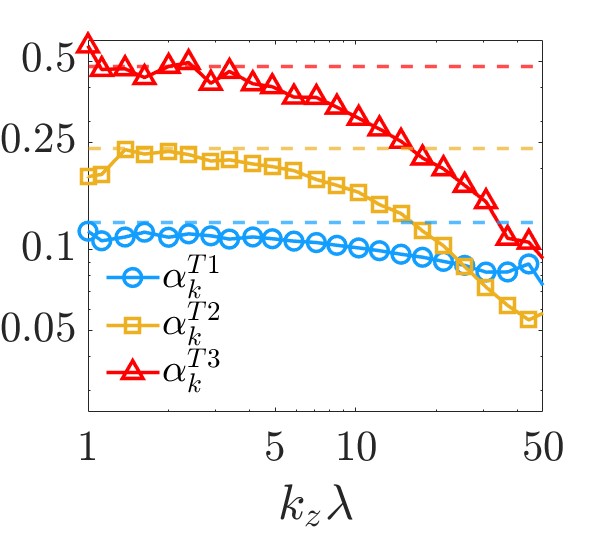}}
  \caption{(a-c): measured damping rate $\tilde{\sigma}_k$ (in red) and dispersion relation $\tilde{\omega}_k$ (in red) as a function of wavenumber $k_z$, in the FOUCAULT model for 
  temperatures $T=1.7, \, 1.95, \, 2.1 \, {\rm K}$(a, b, c, respectively). (d): dimensionless damping rate $\alpha_k$ for $T=1.7, \, 1.95, \, 2.1 \, {\rm K}$ 
  corresponding to curves in blue, yellow and red, respectively.}
  \label{fig:Fig6}
\end{figure}

The temperature dependence of the damping rate is summarized in Fig.~\ref{fig:Fig6}. For each temperature, in Fig.~\ref{fig:Fig6} (a-c) we report the measure of $\tilde{\sigma}_k$ and $\tilde{\omega}_k$, while in Fig.~\ref{fig:Fig6} (d) we report the dimensionless damping rate $\alpha_k$ for varying temperature. Our results show that as temperature increases the damping rate of KWs is enhanced due to stronger mutual friction, which increases the energy transfer from the superfluid vortices to the normal fluid (then dissipated into heat), resulting in more rapid KWs attenuation. We also observe  that the scale-dependent deviation from the constant value is also increasing with temperature. 

\section{Conclusions}

In the present work, we have investigated the propagation of KW on quantized vortices and their impact on the dynamics of the normal component of helium~II. We have also observed how normal fluid fluctuations induced by vortices in the normal fluid have an impact on vortices themselves thanks to the recently developed, two-way fully coupled algorithm FOUCAULT \cite{Galantucci2020}, and compared our results with the one-way coupling model elaborated by Schwarz \cite{schwarz-1978}. Our study shows three main results. First, the dispersion relation computed in the FOUCAULT model is strongly temperature dependent, in contrast to the dispersion relation calculated employing the Schwarz model. Second, the dimensionless damping $\alpha_k$ expressing the ratio of mutual friction dissipation to wave propagation results to be scale-dependent when using the two-way coupling, while it shows to be constant with the Schwarz model. These two first results show that the self-consistent modeling included in FOUCAULT reveals novel features regarding wave propagation on quantum vortices which were not predicted by previous models.

In addition, and more importantly, our third result consists in reporting numerical and theoretical evidence that KW on quantized vortices in superfluid helium can induce a coherent response in the normal fluid through oscillations of the local vorticity field. The measured dispersion relation of these normal fluid waves matches that of the KWs on vortices, confirming a strong dynamical coupling between the two components. The frequency and damping of these waves exhibit a clear temperature dependence governed by the mutual friction parameters.

Our findings are relevant for future experiments employing high-resolution tracer visualization techniques \cite{peretti-etal-2023}. From our simulations we estimate that KWs with an initial amplitude $A_0 \sim 0.1\,\mu$m induce a root-mean-square velocity in the normal fluid of order $10^{-1}\,\mu$m/s. Although this value depends on parameters such as temperature, vortex density, and wave amplitude, larger KW amplitudes may occur in experiments, which could enhance the resulting normal-fluid signal and make it accessible to high-resolution tracer visualization techniques. This opens new possibilities for probing KW dynamics experimentally through the correlated motion of the surrounding normal fluid.

\begin{acknowledgments}
G.K. and S.S. acknowledges support
of Agence Nationale de la Recherche through the project
QuantumVIW ANR-23-CE30-0024-02.
The authors thank P.~Z. Stasiak for valuable discussions and technical assistance.
\end{acknowledgments}

\end{document}